\documentclass[aip,jcp,reprint,groupedaddress]{revtex4-2}
\usepackage{amssymb,amsmath}
\usepackage{graphicx}
\usepackage{array}
\usepackage[whole]{bxcjkjatype}
\usepackage[version=4]{mhchem}
\begin{document}
\title{Analytic theory of nonlinearly coupled electrokinetics in nanochannels}
\author{Yuki Uematsu (植松\hspace{2mm}祐輝)$^{1,2}$ }
\email{uematsu@phys.kyutech.ac.jp}
\affiliation{$^1$Department of Physics and Information Technology, Kyushu Institute of Technology, 820-8502 Iizuka, Fukuoka, Japan\\
$^2$PRESTO, Japan Science and Technology Agency, Kawaguchi, Saitama 332-0012, Japan}
\date{\today}
\pacs{}
\begin{abstract}
The nonlinear electrokinetic response of ionic solutions is important in nanofluidics.
However, quantitatively understanding the mechanisms is still a challenging problem because of a lack of analytic approaches.
Here, a general framework for calculating the nonlinear electrokinetic coefficients of strongly confined electrolytes is constructed using a perturbation scheme of the pressure and voltage differences across a nanochannel.
The theory is applied to an electrically neutral nanochannel filled with electrolytes, and analytic expressions for the first- and third-order electrokinetic coefficients are obtained. 
We demonstrate that the combination of high hydrodynamic permeability and ion-wall friction plays an essential role in nonlinear electrokinetics.
Furthermore, we analytically demonstrate that the external flow induces uniform excess charge inside the nanochannel. 
\end{abstract}

\maketitle

\section{Introduction}
Recent developments in the technology to manipulate, manufacture, and fabricate atoms and molecules at the nanoscale level enable the production of nanofluidic devices that exhibit excellent transport and electrokinetic properties. \cite{Kim2007, Bocquet2010, Majumder2005, Holt2006, Bocquet2016, Esfandiar_2017, Timothee2019, Secchi2016, Poggioli_2019, Alice2020, Marion2020, Noy2021, Koyama_2021, Feng2022, Boon_2022, You_2022}
One of the examples is the ultra-high hydrodynamic permeability of water flowing through carbon nanotubes. \cite{Majumder2005, Holt2006, Bocquet2016}
The carbon surface was experimentally demonstrated to be hydrophobic, with the slip length on the micrometer scale. \cite{Majumder2005, Holt2006, Bocquet2016} 
A large zeta potential \cite{Timothee2019} and anomalous conductance \cite{Secchi2016} of electrolytes confined in nanochannels are also useful for achieving energy conversion at high efficiency.
In addition to these linear responses, electrolyte solutions in strong confinements, typically sub- or few nanometers, exhibit unique nonlinear electrokinetics. \cite{Timothee2019, Poggioli_2019, Alice2020, Boon_2022}   
In strong confinements, the streaming current depends on the applied voltage and the applied pressure difference.
More precisely, the ionic current response has a component proportional to $(\Delta p)(\Delta\psi)^2$ and $(\Delta p)^2(\Delta\psi)$, where $\Delta p$ is the applied pressure difference, and $\Delta\psi$ is the applied voltage difference  \cite{Timothee2019, Alice2020}.
This unique non-linearity is essentially different from the previously reported nonlinear electrokinetics \cite{Bazant2004, Keyser2019, Khair_2022} because in previous studies, the focus was on the electro-osmotic/phoretic velocity proportional to $(\Delta \psi)^2$ and $(\Delta \psi)^3$, where $\Delta p$ and $\Delta \psi$ are not nonlinearly coupled.
Janus particles exhibit a quadratic dependence of the electrokinetic flow  $\sim(\Delta\psi)^2$ because the induced charge on the particle surface is linear with $\Delta\psi$. \cite{Squires2006}
The cubic dependence of the electrophoretic mobility is a result of strong surface conduction in the spherical geometry. \cite{Yariv2014, Keyser2019}
 
Phenomenological nonlinear polynomials for electrokinetic flux were first suggested for glass membranes, ion-exchange membranes, and animal membranes. \cite{Rastogi1993}
However, these experimental data exhibit complex dependence rather than simple polynomial nonlinear dependence of the applied pressure and voltage. 
Nevertheless, the recent nanofluidic measurements on the streaming current \cite{Timothee2019,Alice2020} excellently fit parabolic curves represented by $\sim (\Delta p)^2$ or $\sim (\Delta \psi)^2$, implying that there should exist a novel fundamental physical mechanism of the nonlinear responses.  
At present, numerical simulation of electro-hydrodynamic equations including high hydrodynamic permeability, ion-wall friction, and energy barriers inside the nanochannels can reproduce the quadratic responses of the streaming current on the applied pressure or the voltage, \cite{Timothee2019,Alice2020} but the analytic calculation of each nonlinear electrokinetic coefficient is still lacking.

Here, we construct a general framework to solve electrohydrodynamic equation by perturbation theory with the applied pressure and voltage difference.
By applying this framework to a confined electrolyte solution in an uncharged nanochannel, the linear and nonlinear electrokinetic coefficients are analytically calculated.      
The obtained analytic expression has a simple form in the limit of a long channel, and these equations demonstrate that the combination of the high hydrodynamic permeability and ion-wall friction is the necessary condition for experimentally measurable nonlinear responses.
Furthermore, we investigate how the electrostatic potential profiles along the channel coordinate respond to the perturbations in the applied voltage and water flow. 
Calculation of the induced charge density demonstrates that water flow induces uniform effective charges that are quadratic with the applied voltage and the water flow. 
The flow-induced charge inside the nanochannel is a crucial factor of the third-order electrokinetics in our model.
In the end, we discuss the application of the theory to the previous experiments. \cite{Alice2020} 
Although all of the quantities in the theory cannot be determined by the limited experimental data, the single carbon nanotube exhibits the dimensionless nonlinear electrokinetic coefficients in the order of $10^3$ to $10^5$, which is significantly large. 
Such large nonlinear electrokinetics suggests the combination of large hydrodynamic permeability and ion-wall friction in the nanochannels.
Furthermore, the joint length between the nanochannel and reservoirs is found to be important to quantifying the nonlinear electrokinetic coefficients, whereas the experimental definition of the joint length has not been investigated in the previous nanochannel experiments.  

\begin{figure}[t]
\center
\includegraphics[width = 8.0cm]{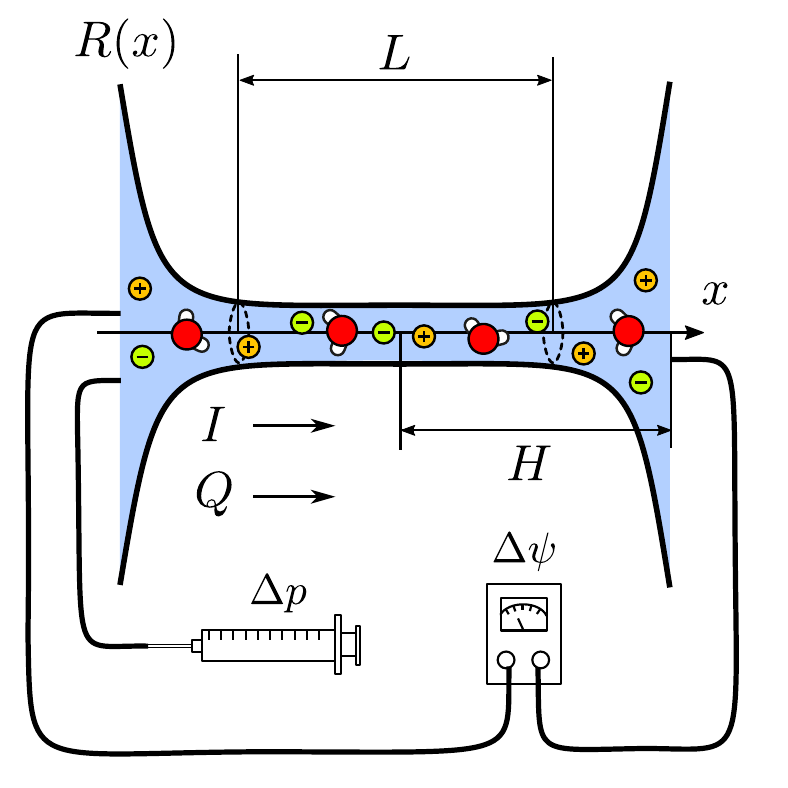}
\caption{
Nanochannel connected with reservoirs filled with monovalent electrolyte solution.
The pressure difference, $\Delta p$, and the voltage difference, $\Delta\psi$, can be externally applied, whereas the solution flux, $Q$, and the ionic current, $I$, are measurable. 
}
\label{fig:1}
\end{figure}

\section{Model}
\subsection{1-dimensional electrohydrodynamic equations}
The model is similar to those suggested to consider the confinement effect in nanochannels \cite{Timothee2019,Alice2020}.
We consider an uncharged nanochannel connected with two reservoirs filled with electrolyte solutions as illustrated in Fig.~\ref{fig:1}. 
The system is assumed to be axisymmetric with respect to the $x$-axis, and the surface of the nanochannel and reservoirs is given by $R(x)$.
In the nanochannel region, $R(x)=R_0$ is constant, where $R_0$ is the radius of the nanochannel, and $L$ is the length of the nanochannel. 
The electrolyte is monovalent and fully dissociated in the solution with the dielectric constant $\varepsilon$ and the viscosity $\eta$.
The electrolyte concentration in the reservoirs is set at $\rho_0$, and the system is stored at the temperature $T$.
The cations and anions are assumed to have the same diffusion constant $D$ in the reservoirs and the nanochannel, but we can easily release this assumption.
The inhomogeneity of the ion concentration, the electrostatic potential, and ion and water velocity fields in the radial and angular coordinates are neglected, and their inhomogeneity is only considered in the axial direction.
Furthermore, the ionic volume interactions \cite{Kilic_2007} are neglected even though the salt concentration is high, and anisotropy and reduction of the dielectric constant inside the nanochannel \cite{Loche_2019} are neglected.
In the nanotube, the ions feel the energy barrier due to the different solvation structure from bulk, but this is omitted because we consider the high salinity where the electrostatic screening reduces the energy barrier \cite{Fornasiero_2008}.
These simplifications yield a set of electro-hydrodynamic equations 
for the pressure field $p(x)$, the water velocity field $v_\mathrm{w}(x)$, the ion velocity field $v_i(x)$ ($i=\pm$ denotes cation and anion), the $i$ type ion concentration $\rho_i(x)$, and the electrostatic potential $\psi(x)$ given as follows:
\begin{eqnarray}
	-\frac{d}{dx} \left[\pi R(x)^2 v_\mathrm{w}(x)\right] =0, &&\label{eq.20}\\
	-\frac{d}{dx} \left[\pi R(x)^2 \rho_i(x)v_i(x)\right] =0,&& \label{eq.21}\\
-\frac{dp(x)}{dx} -\sum_{i=\pm}\rho_i(x)\xi_i(x)[v_\mathrm{w}(x)-v_i(x)] \qquad && \nonumber \\
	- \frac{2\lambda_0(x)}{R(x)} v_\mathrm{w}(x) =0, && \label{eq.22}\\
	-\frac{d\mu_i}{dx}-\xi_i(x)[v_i(x)-v_\mathrm{w}(x)]-\lambda_i(x)v_i(x) = 0, &&\label{eq.23}\\
	\frac{1}{R(x)^2}\frac{d}{dx} \left[\varepsilon \varepsilon_0 R(x)^2 \frac{d}{dx}\psi(x)\right] =-e\sum_{i=\pm}q_i\rho_i(x), && \label{eq.24}
\end{eqnarray}
where $\xi_i(x)$ is the friction coefficient between ion and water, $\lambda_0(x)$ is the friction coefficient between water and wall per surface area, $\lambda_i(x)$ is the friction coefficient between ion and wall, $\varepsilon_0$ is the vacuum permittivity, $e$ is the elementary charge, and $q_i(=\pm 1)$ is the valence of the ions. 
The electrochemical potential of ions, $\mu_i$ is defined by  
\begin{equation}
\mu_i(x)=k_\mathrm{B}T\ln\frac{\rho_i(x)}{\rho_0}+eq_i\psi(x),
\end{equation}
where $k_\mathrm{B}$ is the Boltzmann constant.
Eqs. \ref{eq.20} and \ref{eq.21} are the continuity equations for the water flux and the ionic flux.
Eqs.~\ref{eq.22} and \ref{eq.23} are the force balance equations for water and ions, whereas eq.~\ref{eq.24} is the Poisson equation.
A similar model was first introduced for explaining nonlinear electrokinetics in the subnanometer channels \cite{Timothee2019} and also in a single carbon nanotube. \cite{Alice2020}
Eqs.~\ref{eq.22} and \ref{eq.23} can be derived by the Onsager variational principle \cite{Doi2011} as described in supplementary material.

For analytical tractability, we introduce the profile of the ionic diffusion constant $D_i(x)$ and the profile of the ionic friction ratio $\beta_i(x)$ as follows:
\begin{eqnarray}
D_i(x)	   &=&\frac{k_\mathrm{B}T}{\xi_i(x)+\lambda_i(x)},\\
\beta_i(x) &=&\frac{\xi_i(x)}{\xi_i(x)+\lambda_i(x)}.
\end{eqnarray}
Furthermore, in this study, we consider a specific geometry $R(x)$ explicitly given by:
\begin{equation}
R(x)=\left\{\begin{array}{lcc}
\displaystyle {R_0}&\mathrm{for}&|x|<\displaystyle\frac{L}{2}\\\\
\displaystyle {R_0}\mathrm{e}^\frac{|x|-L/2}{2\delta}&\mathrm{for}&|x|>\displaystyle \frac{L}{2}\\
\end{array},\right.\label{eq:R}
\end{equation}
where $R_0$ is the radius of the nanochannel, $L$ is the nanochannel length and $\delta$ is the characteristic length at the joint between the nanochannel and reservoirs. 
In addition, we use a sharp step function for $\lambda_0(x)$ and $\beta_i(x)$ given by:
\begin{eqnarray}
\lambda_0(x) & = & \frac{\eta}{b}\theta(\frac{L}{2}-|x|),\\\nonumber\\
\beta_i(x)   & = & \alpha_i + (1-\alpha_i)\theta(|x|-\frac{L}{2}),
\end{eqnarray} 
where $\eta$ is the solution viscosity, $b$ is the slip length of the nanochannel, and $\alpha_i$ is the ion-wall friction ratio.
The case of $\alpha_i=1$ corresponds to no ion-wall friction force, and the case of $\alpha_i=0$ corresponds to only ion-wall friction (no ion-water friction).
Furthermore, we use the homogeneous diffusion profile $D_+(x)=D_-(x)=D$ for both cations and anions. 
This limitation can be easily released in the following analytical calculation.
The boundary conditions of the equations are: 
\begin{eqnarray}
p(-H) &=& -p(H)=\frac{\Delta p}{2},\label{eq.31}\\
\psi(-H) &=& -\psi(H)=\frac{\Delta \psi}{2},\label{eq.32}\\
\rho_i(-H) &=& \rho_i(H)=\rho_0,\label{eq.33}
\end{eqnarray}
where $H$ is half of the system size, $\Delta p$ is the applied pressure difference, and $\Delta\psi$ is the applied voltage.

\subsection{Derivation of the integro-differential Poisson equation}
In this section, we derive an integro-differential Poisson equation, eq.~\ref{eq.a27} from eqs.~\ref{eq.20} to \ref{eq.24}.
The similar method was first introduced in Ref. \citenum{Alice2020}. 
Eqs.~\ref{eq.20} and \ref{eq.21} are directly solved by introducing the three integral constants: the water flux, $Q$, and the $i$ type ionic current, $J_i$.
The water velocity and ionic currents are solved as:
\begin{eqnarray}
v_\mathrm{w}(x) &=& \frac{Q}{\pi R^2(x)},\label{eq.29}\\
\rho_i(x)v_i(x) &=& \frac{J_i}{\pi R^2(x)}\label{eq.30}.
\end{eqnarray}
At first, we introduce the effective potential, $W_i(x)$, and the effective concentration, $c_i(x)$, as
\begin{eqnarray}
W_i(x)&=&q_i\varphi(x)-\mathrm{Pe}\frac{\pi R_0^2}{L}\int^x_0\frac{\beta_i(x')}{\pi R(x')^2}dx',\label{eq.a1}\\
c_i(x)&=& \frac{\rho_i(x)}{\rho_0}\mathrm{e}^{W_i(x)},\label{eq.a2}
\end{eqnarray}
where the P\'eclet number $\mathrm{Pe}$ is defined by:
\begin{equation}
	\mathrm{Pe}=\frac{QL}{\pi R_0^2D}. \label{eq:Pe}
\end{equation}
Solving eq.~\ref{eq.23} with respect to the ionic velocity $v_i$, and plugging eqs.~\ref{eq.a1} and \ref{eq.a2} into $v_i$, we obtain: 
\begin{equation}
v_i(x) =  -\frac{1}{\xi_i(x)+\lambda_i(x)}\frac{d\mu_i}{dx}+\beta_i(x)v_\mathrm{w}(x)=-\frac{D}{c_i(x)}\frac{d c_i}{dx}.\label{eq.a3} 
\end{equation}
Because $J_i=\pi R^2(x)\rho_i(x)v_i(x)=\textrm{const.}$, we obtain:
\begin{equation}
\frac{d c_i}{dx} = -\frac{J_i}{\rho_0 D}\frac{\mathrm{e}^{W_i(x)}}{ \pi R(x)^2}.
\label{eq.a4}
\end{equation}
Eq.~\ref{eq.a4} can be solved with the boundary conditions $\rho_i(-H)=\rho_0$ and $\varphi(-H)=\Delta\varphi/2$, as: 
\begin{equation}
c_i(x)=\mathrm{e}^{\frac{q_i\Delta\varphi+p_i\mathrm{Pe}}{2}}-\frac{J_i}{\rho_0 D}f_i(x),
\label{eq.a5}
\end{equation} 
where
\begin{eqnarray}
p_i &=& \frac{\pi R_0^2}{L}\int^{H}_{-H}\frac{\beta_i (x') dx'}{\pi R(x')^2},\label{eq.a6}\\ 
f_i(x)&=&\int^x_{-H}\frac{\mathrm{e}^{W_i(x')}}{\pi R(x')^2}dx'.\label{eq.a24}
\end{eqnarray}
The boundary conditions, $\rho_i(H)=\rho_0$ and $\varphi(H)=-\Delta\varphi/2$, yield
\begin{equation}
c(H)=\mathrm{e}^{W_i(H)}=\mathrm{e}^{-(q_i\Delta\varphi+p_i\mathrm{Pe})/2},
\label{eq.a8}
\end{equation}
and solving eqs.~\ref{eq.a5} and \ref{eq.a8} with respect to $J_i$ gives
\begin{equation}
J_i=\frac{2\rho_0 D}{f_i(H)}\sinh\frac{q_i\Delta\varphi+p_i\mathrm{Pe}}{2}\label{eq.a9}.
\end{equation}
Therefore, eqs.~\ref{eq.20} to \ref{eq.24} can be converted into a single integro-differential Poisson equation rewritten as:
\begin{equation}
\frac{1}{\kappa^2R(x)^2}\frac{d}{dx}\left[R(x)^2\frac{d}{dx}\right]\varphi=-\frac{1}{2}\sum_{i=\pm}q_ic_i(x)\mathrm{e}^{-W_i(x)},\label{eq.a27}
\end{equation}
where $\kappa^2 = 2e^2\rho_0/\varepsilon\varepsilon_0 k_\mathrm{B}T$, $c_i(x)$, $f_i(x)$, and $W_i(x)$ are defined by eqs.~\ref{eq.a1}, \ref{eq.a5}, \ref{eq.a24}, and \ref{eq.a9}.

For given $Q$ (or $\mathrm{Pe}$) and $\Delta\varphi$, eq.~\ref{eq.a27} yields the electrostatic potential profile $\varphi(x)$.
The individual ionic current, $J_i$, can be calculated by eqs. \ref{eq.a24} and \ref{eq.a9}, and the ionic current is given by:
\begin{equation}
I=e \sum_iq_i  J_i.
\label{eq:I}
\end{equation}  
The pressure distribution can be calculated by integrating eq.~\ref{eq.22} as:
\begin{equation}
\begin{split}
	p(x) = &-\int^x_0 \frac{2\lambda_0(x')}{R(x')}\frac{Q}{\pi R(x')^2}dx' \\
	& -\frac{k_\mathrm{B}T}{D}\int^x_0 \frac{\sum_i\beta_i(x')\left(\rho_i(x')Q-J_i\right)}{\pi R(x')^2}dx',
\label{eq.a13}
\end{split}
\end{equation}
and the pressure difference is given by $\Delta p = p(-H)-p(H)$.

\section{Perturbation theory}

The applied voltage and pressure difference, $\Delta \psi$ and $\Delta p$, generate the ionic current $I$ and the solution flux $Q$ across the nanochannel.
We consider the perturbation expansion of the fluxes as functions of the fields given by: 
\begin{eqnarray}
Q &=& D \frac{\pi R_0^2}{L} \sum_{n+m=0}^\infty Q^n_m\frac{(\Delta P)^n}{n!}\frac{(\Delta\varphi)^m}{m!},\label{eq1}\\
I &=& e\rho_0 D \frac{\pi R_0^2}{L}\sum_{n+m=0}^\infty I^n_m\frac{(\Delta P)^n}{n!}\frac{(\Delta\varphi)^m}{m!},\label{eq2}
\end{eqnarray}
where $\Delta P=\Delta p/\rho_0 k_\mathrm{B}T$ and $\Delta \varphi = e\Delta\psi/k_\mathrm{B}T$ are the dimensionless pressure difference and applied voltage, respectively. 
The summation $\sum_{n+m=0}^\infty$ denotes the sum of the terms $(n,m)=(0,0),(1,0),(0,1),(2,0),(1,1),(0,2),\cdots$. 
$Q_0^0$ and $I^0_0$ are zero by definition. 
$Q^1_0$, $Q^0_1$, $I^1_0$, and $I^0_1$ are the linear electrokinetic coefficients, whereas $Q^n_m$ and $I^n_m$ for $n+m\ge 2$ are the nonlinear electrokinetic coefficients.
In this study, we obtain the analytic expressions of the electrokinetic coefficients up to the order of $n+m=3$.
Detailed derivation of the framework of the perturbation theory is explained in the following sections. 

\subsection{Derivation of expanded Poisson equations}
In this section, we derive expanded Poisson equations of each order $(n,m)$. 
To solve eq.~\ref{eq.a27} by the perturbation theory, it is necessary to expand eq.~\ref{eq.a27} with respect to $\mathrm{Pe}$ and $\Delta\varphi$ not $\Delta P$ and $\Delta \varphi$.
At first, we expand $\varphi(x)$,  $W_i(x)$, and  $c_i(x)$ with respect to $\mathrm{Pe}$ and $\Delta\varphi$,  
\begin{eqnarray}
\varphi(x) &=& \sum_{n+m=0}^\infty{\varphi}^n_m(x)\frac{(\mathrm{Pe})^n}{n!}\frac{(\Delta\varphi)^m}{m!},\\ 
W_i(x)&=& \sum_{n+m=0}^\infty{W_i}^n_m(x)\frac{(\mathrm{Pe})^n}{n!}\frac{(\Delta\varphi)^m}{m!},\\
c_i(x)&=& \sum_{n+m=0}^\infty{c_i}^n_m(x)\frac{(\mathrm{Pe})^n}{n!}\frac{(\Delta\varphi)^m}{m!}.
\end{eqnarray}
Using eq.\ref{eq.a1}, the expanded ${W_i}^n_m(x)$ is given by:
\begin{equation}
{W_i}^n_m(x)=\left\{\begin{array}{ll}
q_i\varphi^n_m(x)& \textrm{for }(n,m)\neq(1,0),\\\\
\displaystyle q_i\varphi^1_0(x)-\frac{\pi R_0^2}{L} & \displaystyle  \int^x_0 \frac{\beta_i(x')}{\pi R(x')^2}dx'\\\\
& \textrm{for }(n,m)=(1,0).
\end{array}\right.
\end{equation}
To calculate ${c_i}^n_m(x)$, we need to expand ${f_i}(x)$ as:
\begin{equation}
f_i(x)= \frac{L}{\pi {R_0}^2}\sum_{n+m=0}^\infty{f_i}^n_m(x)\frac{(\mathrm{Pe})^n}{n!}\frac{(\Delta\varphi)^m}{m!}.
\end{equation}
Then, we obtain the expansion for ${f_i}(x)/{f_i}(H)$,
\begin{equation}
\frac{f_i(x)}{f_i(H)}=\frac{{f_i}^0_0(x)}{{f_i}^0_0(H)}+\frac{{f_i}^1_0(x)}{{f_i}^0_0(H)}\mathrm{Pe}+\frac{{f_i}^0_1(x)}{{f_i}^0_0(H)}\Delta\varphi+\cdots,
\end{equation}
where we use ${f_i}^1_0(H)={f_i}^0_1(H)=0$. 
Then, we obtain ${c_i}^n_m(x)$ up to the second order $n+m=2$, 
\begin{eqnarray}
{c_i}^0_0(x)&=&1,\\
{c_i}^1_0(x)&=&p_i\left(\frac{1}{2}-h{f_i}^0_0(x)\right),\\
{c_i}^0_1(x)&=&q_i\left(\frac{1}{2}-h{f_i}^0_0(x)\right),\\
{c_i}^2_0(x)&=&\frac{p_i^2}{4}-2p_ih{f_i}^1_0(x),\\
{c_i}^1_1(x)
&=&\frac{p_iq_i}{4}-q_ih{f_i}^1_0(x)-p_ih{f_i}^0_1(x),\\
{c_i}^0_2(x)
&=&\frac{1}{4}-2q_ih{f_i}^0_1(x).
\end{eqnarray}
where $h=1/{f_i}^0_0(H)$ is $i$-independent. 
For convenience of the following analysis, we define a functional $F[x;g(x)]$ by:
\begin{equation}
F[x;g(x)] = \frac{\pi {R_0}^2}{L}\int^{x}_{-H}\frac{g(x')dx'}{\pi R(x')^2}.
\label{eq:F}
\end{equation}
Then, we obtain the expressions for ${f_i}^n_m(x)$ up to the second order given by:
\begin{eqnarray}
{f_i}^0_0(x)&=&F(x;1),\label{eq:fi00}\\
{f_i}^1_0(x)&=&F(x;{W_i}^1_0(x)),\label{eq:fi10}\\
{f_i}^0_1(x)&=&F(x;{W_i}^0_1(x)),\label{eq:fi01}\\
{f_i}^2_0(x)&=&F(x;{W_i}^2_0(x)+{W_i}^1_0(x)^2),\label{eq:fi20}\\
{f_i}^1_1(x)&=&F(x;{W_i}^1_1(x)+{W_i}^1_0(x){W_i}^0_1(x)),\label{eq:fi11}\\
{f_i}^0_2(x)&=&F(x;{W_i}^0_2(x)+{W_i}^0_1(x)^2),\label{eq:fi02}
\end{eqnarray}
where we use ${W_i}^0_0(x)=0$.
For the zeroth order, ${c_i}^0_0(x)=1$ and ${W_i}^0_0(x)=q_i\varphi^0_0(x)$ give the zeroth-order Poisson equation (usual Poisson-Boltzmann equation),
\begin{equation}
\frac{1}{\kappa^2 R^2(x)}\frac{d}{dx}\left[R^2(x)\frac{d\varphi^0_0(x)}{dx}\right]=\sinh\varphi^0_0,
\label{eq.b7}
\end{equation}
with the boundary condition $\varphi^0_0(-H)=\varphi^0_0(H)=0$.
The solution of eq.~\ref{eq.b7} is $\varphi^0_0(x)=0$.
The higher order expanded Poisson equations are given by:
\begin{equation}
\frac{1}{\kappa^2 R^2(x)}\frac{d}{dx}\left[R^2(x)\frac{d\varphi^n_m(x)}{dx}\right] = -\rho^n_m(x),
	\label{eq:PBhigh}
\end{equation} 
where $\rho^n_m(x)=\frac{1}{2}\sum_i q_i{\rho_i}^n_m(x)$ is induced charge density, and the expressions for ${\rho_i}^n_m(x)$ up to the second order are given by: 
\begin{eqnarray}
{\rho_i}^1_0(x) &=& {c_i}^1_0(x)-{W_i}^1_0(x),\label{eq.b8}\\
{\rho_i}^0_1(x) &=&{c_i}^0_1(x)-{W_i}^0_1(x),\label{eq.b9}\\
{\rho_i}^2_0(x)&=&{c_i}^2_0(x)-2{c_i}^1_0(x){W_i}^1_0(x)-{W_i}^2_0(x)+{W_i}^1_0(x)^2,\label{eq.b10}\nonumber\\\\
{\rho_i}^1_1(x)&=&{c_i}^1_1(x)-{c_i}^1_0(x){W_i}^0_1(x)-{c_i}^0_1(x){W_i}^1_0(x)\nonumber\\
&&-{W_i}^1_1(x)+{W_i}^0_1(x){W_i}^1_0(x),\label{eq.b11}\\
{\rho_i}^0_2(x)&=&{c_i}^0_2(x)-2{c_i}^0_1(x){W_i}^0_1(x)-{W_i}^0_2(x)+{W_i}^0_1(x)^2\label{eq.b12}.\nonumber\\
\end{eqnarray}
In the derivation of eqs.~\ref{eq.b8} to \ref{eq.b12}, we use ${c_i}^0_0(x)=1$, ${W_i}^0_0(x)=0$, and
\begin{equation}
\begin{split}
\mathrm{e}^{\pm W_i(x)}& = 1\pm {W_i}^1_0(x)\mathrm{Pe}\pm{W_i}^0_1(x)\Delta\varphi\\
&+[{W_i}^1_0(x)^2\pm {W_i}^2_0(x)]\frac{(\mathrm{Pe})^2}{2}\\
&+[{W_i}^1_0(x){W_i}^0_1(x)\pm{W_i}^1_1(x)](\mathrm{Pe})(\Delta\varphi)\\
&+[{W_i}^0_1(x)^2\pm{W_i}^0_2(x)]\frac{(\Delta\varphi)^2}{2}+\cdots.
\end{split}
\end{equation}
The boundary conditions of eq.~\ref{eq:PBhigh} are $\varphi^n_m(-H)=\varphi^n_m(H)=0$ for $(n,m)\neq(0,1)$ and $\varphi^0_1(-H)=-\varphi^0_1(H)=1/2$ for $(n,m)=(0,1)$.
Because $\rho^n_m(x)$ ($n+m\ge 1$) has a linear term $-\varphi^n_m(x)$, we can extract the source term of the linear differential equations by $S^n_m(x)=\rho^n_m(x)+\varphi^n_m(x)$ which are given by:  
\begin{eqnarray}
&&S^1_0(x)=\frac{1}{2}\nonumber\\
&&\times\left[{c_+}^1_0(x)-{c_-}^1_0(x)+\frac{\pi {R_0}^2}{L}\int^x_0 \frac{\beta_+(x')-\beta_-(x')}{\pi R(x')^2}dx'\right],\nonumber\\\\
&&S^0_1(x)=\frac{1}{2}\left[{c_+}^0_1(x)-{c_-}^0_1(x)\right],\\
&&S^2_0(x)=\sum_{i=\pm}\frac{q_i}{2}\left[{c_i}^2_0(x)+{W_i}^1_0(x)^2-2{c_i}^1_0(x){W_i}^1_0(x)\right],\nonumber\\\\
&&S^1_1(x)=\sum_{i=\pm}\frac{q_i}{2}\left[{c_i}^1_1(x)+{W_i}^1_0(x){W_i}^0_1(x)\right.\nonumber\\
&&\qquad\qquad\qquad\qquad\left.-{c_i}^1_0(x){W_i}^0_1(x)-{c_i}^0_1(x){W_i}^1_0(x)\right],\\
&&S^0_2(x)=\sum_{i=\pm}\frac{q_i}{2}\left[{c_i}^0_2(x)+{W_i}^0_1(x)^2-2{c_i}^0_1(x){W_i}^0_1(x)\right].\nonumber\\
\end{eqnarray}

\subsection{Solution of expanded Poisson equation in each order}
The first-order and the second-order Poisson equations are obtained as:
\begin{equation}
\frac{1}{\kappa^2 R^2(x)}\frac{d}{dx}\left[R^2(x)\frac{d\varphi^n_m(x)}{dx}\right] - \varphi^n_m(x) = -S^n_m(x),
\end{equation} 
Further calculation yields analytical expressions for $S^1_0(x)$ and $S^0_1(x)$ as follows:
\begin{eqnarray}
S^1_0(x)&=&\left\{\begin{array}{ll}
\displaystyle -\mathrm{sgn}(x)\left[-\frac{\sigma_0}{2}h\nu t(x)\right], & \displaystyle \textrm{for } |x|>\frac{L}{2}\\\\
\displaystyle \sigma_0 h\nu t'\frac{x}{L}, & \displaystyle \textrm{for } |x|<\frac{L}{2}
\end{array},\right.\label{eq.S10}\\\nonumber\\
S^0_1(x)&=&\left\{\begin{array}{ll}
\displaystyle -\mathrm{sgn}(x)\left[\frac{1}{2}-h\nu t(x)\right], & \displaystyle \textrm{for } |x|>\frac{L}{2}\\\\
\displaystyle -h\frac{x}{L}, & \displaystyle \textrm{for } |x| < \frac{L}{2}
\end{array},\right.
\end{eqnarray}
where $\sigma_0  = \sum_i q_i\alpha_i$, $\nu = \delta/L$, $t(x) = \mathrm{e}^{-\frac{|x|-L/2}{\delta}}-\mathrm{e}^{\frac{H-L/2}{\delta}}$, and $t'=t(\frac{L}{2})$.
Using this notation, we have $h=1/(1+2\nu t')$. 
For $(n,m)=(1,0)$ and $(0,1)$ we obtain the approximated solutions,
\begin{eqnarray}
\varphi^1_0(x)&=&S^1_0(x),\label{eq:phi10}\\
\varphi^0_1(x)&=&S^0_1(x).\label{eq:phi01}
\end{eqnarray}
The exact solution is given by supplementary material.

Using the approximate solutions for $\varphi^1_0(x)$ and $\varphi^0_1(x)$, the analytical expressions for $S^2_0(x)$, $S^1_1(x)$, and $S^0_2(x)$ are obtained as follows:
\begin{eqnarray}
S^2_0(x) &=& \left\{\begin{array}{ll}
\displaystyle \frac{\sigma_0 (2-\sigma_1)}{4}h^2\nu^2t(x)^2, & \displaystyle \textrm{for } |x|>\frac{L}{2},\\\\
\displaystyle \frac{\sigma_0 (2-\sigma_1)}{8} h\nu t' -\frac{\sigma_0 (2-\sigma_1)}{2}&\displaystyle h^2\nu t'\frac{x^2}{L^2},\\
& \displaystyle \textrm{for } |x|<\frac{L}{2},
\end{array}\right.\label{eq:S66}\\\nonumber\\\nonumber\\
S^1_1(x)&=&\left\{\begin{array}{ll}
\displaystyle \frac{2-\sigma_1}{4}h^2\nu^2t(x)^2, & \displaystyle \textrm{for } |x|>\frac{L}{2},\\\\
\displaystyle \frac{2-\sigma_1}{8}h\nu t'-\frac{2-\sigma_1}{2}h^2\nu t'\frac{x^2}{L^2}, & \displaystyle \textrm{for } |x|<\frac{L}{2},
\end{array}\right.\label{eq:S67}\\\nonumber\\\nonumber\\
S^0_2(x) &= &0,
\end{eqnarray}
where $\sigma_1 = \sum_i \alpha_i$. 
For $(n,m)=(2,0)$,$(1,1)$, and $(0,2)$, we obtain the approximated solutions,
\begin{eqnarray}
\varphi^2_0(x)&=& S^2_0(x),\label{eq:75}\\
\varphi^1_1(x)&=& S^1_1(x),\label{eq:76}\\
	\varphi^0_2(x) &=&0.\label{eq:74}
\end{eqnarray}
Derivation of the exact solution is described in supplementary material.
Using eqs.~\ref{eq.S10}--\ref{eq:74}, the induced charge density $\rho^n_m(x)$ of each order inside the nanochannel is obtained as: 
\begin{eqnarray}
\rho^1_0(x) &=& 0,\\
\rho^0_1(x) &=& 0,\\
\rho^2_0(x) &=& \frac{\sigma_0(2-\sigma_1)}{\kappa^2}h^2\nu t'\frac{1}{L^2}, \label{eq:79}\\
\rho^1_1(x) &=& \frac{2-\sigma_1}{\kappa^2}h^2\nu t' \frac{1}{L^2},\label{eq:80}\\
\rho^0_2(x) &=& 0.
\end{eqnarray} 

After obtaining $\varphi^n_m(x)$ and $\rho^n_m(x)$ up to the second order, we can calculate $\Delta p^n_m$ and ${J_i}^n_m$ up to the third order.
The individual ionic current $J_i$ defined in eq.~\ref{eq.a9} is expanded with respect to $\mathrm{Pe}$ and $\Delta\varphi$ as:
\begin{equation}
J_i(\mathrm{Pe},\Delta\varphi)=\rho_0 D\frac{\pi R_0^2}{L}\sum_{n+m=0}^\infty{J_i}^n_m\frac{(\mathrm{Pe})^n}{n!}\frac{(\Delta\varphi)^m}{m!}.\label{eq:Ji}
\end{equation}
We also calculate $\Delta p^n_m$ which is defined by:
\begin{equation}
\Delta p (\mathrm{Pe},\Delta\varphi)= \rho_0 k_\mathrm{B}T\sum_{n+m=0}^\infty \Delta p^n_m \frac{(\mathrm{Pe})^n}{n!}\frac{(\Delta\varphi)^m}{m!}.
\end{equation}
Explicit formulae for ${J_i}^n_m$ and $\Delta p^n_m$ are available in supplementary material.

Because our framework of the perturbation theory is based on the expansion with $\mathrm{Pe}$ and $\Delta\varphi$, we need to change the variables into $\Delta P$ and $\Delta\varphi$. 
The detailed procedure and explicit formulae for $Q^n_m$ and $I^n_m$ are given in supplementary material.

\section{Results and Discussion}
\begin{figure}[t]
\includegraphics[width = 8.6cm]{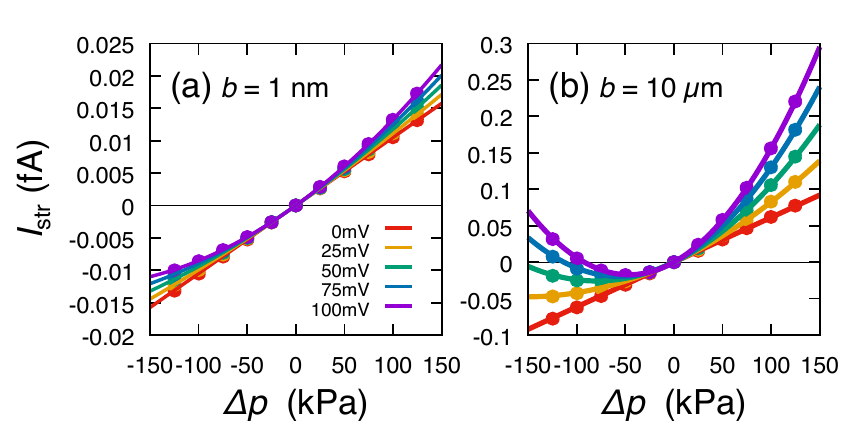}
\caption{
	The voltage-dependent streaming currents $I_\mathrm{str}(\Delta p, \Delta\psi)$ are plotted for (a) low hydrodynamic permeability $b=1\,$nm and (b) high hydrodynamic permeability $b=10\,\mu$m. 
The points are numerically obtained by solving eq.~\ref{eq.a27}, whereas the lines are eq.~\ref{eq2} with analytically obtained coefficients up to $n+m=3$.
The different color represents $\Delta\psi=0$, $25$, $50$, $75$, and $100\,$mV. 
}
\label{fig:2}
\end{figure}

In this section, we investigate typical parameter sets of nanochannels and compare full numerical solutions with analytical results of the perturbation theory. 
Fig.~\ref{fig:2} shows the voltage-dependent streaming current $I_\mathrm{str}(\Delta p, \Delta\psi)=I(\Delta p, \Delta\psi) - I(0, \Delta\psi)$ which was recently measured in the experiments. \cite{Timothee2019, Alice2020}
The ion-wall frictions are set at $\alpha_+=0.002$ and $\alpha_-=0.001$. 
Other parameters are set at $L=1\,\mu$m, $R_0=1\,$nm, $D=2\times10^{-9}\,$m$^2$/s, $\eta = 1\,$mPa/s, $T = 300\,$K, $\varepsilon = 80$, $\rho_0 = 1\,$M, $\delta = 250\,$nm, and $2H=1.25\,\mu$m.
The lines are calculated analytically up to the terms of $n+m=3$ of eq.~\ref{eq2}, and the points are obtained by numerically solving the electro-hydrodynamic equations.
The different color represents the different voltage $\Delta\psi = 0$, $25$, $50$, $75$, and $100\,$mV.
The analytical lines perfectly agree with the points obtained numerically, indicating that our analytic theory is correct.
In the low hydrodynamic permeability case (a) $b=1\,$nm, the nonlinear streaming current is not so drastic, whereas in the high hydrodynamic permeability case (b) $b=10\,\mu$m, the streaming current exhibits a quadratic dependence on the applied pressure difference.
Because the numerical solution includes higher-order terms $n+m>3$, these terms are found to be negligible in this condition.

Although the analytical expressions for all of the electrokinetic coefficients up to the third order are complicated (eqs.~S39--S50 in supplementary material), we take a long-channel limit, $\nu=\delta/L\to 0$, and the infinite system size ($2H\to\infty$) to derive simplified analytic expressions.
Leading terms of the linear coefficients, $Q^1_0$, $Q^0_1$, $I^1_0$, and $I^0_1$, are obtained as follows:
\begin{eqnarray}
Q^1_0&=&K,\label{eq.Q10}\\
Q^0_1&=&I^1_0 = K(\alpha_+-\alpha_-),\label{eq.Q01}\\
I^0_1&=&2+K(\alpha_+-\alpha_-)^2,\label{eq.I01}
\end{eqnarray}
where $K(>0)$ is the non-dimensional hydrodynamic permeation coefficient, which is given by:
\begin{equation}
\frac{1}{K}=\mathrm{Pe}_\mathrm{osm}^{-1}+\sum_i\alpha_i(1-\alpha_i),
\end{equation}
and $\mathrm{Pe}_\mathrm{osm}=\rho_0 k_\mathrm{B}T R_0 b/ 2\eta D$ denotes the hydrodynamic permeation coefficient without the effect of ion-wall friction. 
The ion-wall friction modifies the hydrodynamic permeation coefficient from $\mathrm{Pe}_\mathrm{osm}$ to $K$. 
Similarly, the ionic conductance $I^0_1$ is modified from $2$ to $2+K(\alpha_+-\alpha_-)^2$ due to the ion-wall friction.
Because $Q^1_0$ and $I^0_1$ are diagonal components of the macroscopic transport matrix, they are never negative as a result of positive entropy production.
The coefficients of the electro-osmosis and streaming current are the same, $Q^0_1 = I^1_0$, reflecting the Onsager reciprocal relation.
Interestingly, even though the nanochannel is uncharged in this model, the electro-osmotic mobility (or the zeta potential) is non-zero.
When the anions are more bounded to the nanochannel than the cations, {\it i.e.} $\alpha_+>\alpha_-$, it is intuitive that the zeta potential becomes negative due to the difference between the ion-wall frictions.
In experiments, the nanochannel/water interface is considered to be spontaneously charged due to the friction asymmetry as well as the ion adsorption. 

All the second-order coefficients, $Q^2_0$, $Q^1_1$, $Q^0_2$, $I^2_0$, $I^1_1$, and $I^0_2$, are strictly zero because of the symmetric geometry of the nanochannel.
All the third-order coefficients are proportional to $\delta/L$ in the lowest order.
Neglecting the higher order terms $(\delta/L)^2$, the analytical expressions for the water flux are obtained as follows: 
\begin{eqnarray}
Q^3_0&=&-\frac{K^4}{4}\sigma_1\sigma_2(2-\sigma_1)\frac{\delta}{L},\label{eq.Q30}\\
Q^2_1&=&\frac{K^3}{12}\sigma_0\sigma_1(2-\sigma_1)\left(1-3K\sigma_2\right)\frac{\delta}{L},\label{eq.Q21}\\
Q^1_2&=&\frac{K^3}{12}{\sigma_0}^2\sigma_1(2-\sigma_1)\left(2-3K\sigma_2\right)\frac{\delta}{L},\label{eq.Q12}\\
Q^0_3&=&\frac{K^3}{4}{\sigma_0}^3\sigma_1(2-\sigma_1)\left(1-K\sigma_2\right)\frac{\delta}{L},\label{eq.Q03}
\end{eqnarray}
where $\sigma_2=\sum_i\alpha_i(1-\alpha_i)$, and for the ionic current, 
\begin{eqnarray}
I^3_0&=&\frac{K^3}{4}\sigma_0\sigma_1(2-\sigma_1)\left(1-K\sigma_2\right)\frac{\delta}{L}\label{eq.I30},\\
I^2_1&=&\frac{K^2}{12}\sigma_1(2-\sigma_1)\left[2+K{\sigma_0}^2(4-3K\sigma_2)\right]\frac{\delta}{L},\label{eq.I21}\\
I^1_2&=&\frac{K^2}{12}\sigma_0\sigma_1(2-\sigma_1)\left[4+K{\sigma_0}^2(5-3K\sigma_2)\right]\frac{\delta}{L},\label{eq.I12}\\
I^0_3&=&\frac{K^2}{4}{\sigma_0}^2\sigma_1(2-\sigma_1)\left[2+K{\sigma_0}^2(2-K\sigma_2)\right]\frac{\delta}{L}\label{eq.I03}.
\end{eqnarray}
Eqs.~\ref{eq.Q30} to \ref{eq.I03} show that all the third order coefficients are proportional to $2-\sigma_1$, and $K^2$.
This clearly demonstrates that the large hydrodynamic permeation, $K\gg1$, is necessary for strong third-order responses.
Furthermore, zero ion-wall friction ($\sigma_1=\alpha_++\alpha_-=2$) remove all the third-order responses.  
According to the Onsager reciprocal relation for nonlinear transport, \cite{Rastogi1993} the relations: $Q^2_1=I^3_0$, $Q^1_2=I^2_1$, and $Q^0_3=I^1_2$, are expected.
However, these relations do not hold in eqs.~\ref{eq.Q30} to \ref{eq.I03} because our framework includes the responses of the induced charge density for $n+m=2$, which is discussed in the next paragraph. 
In fact, these relation did not hold in the previous membrane experiments, \cite{Rastogi1993} and thus, we think our theory correctly reflects the properties of nonlinear responses.   

\begin{figure}[t]
\begin{center}
\includegraphics[width = 8.6cm]{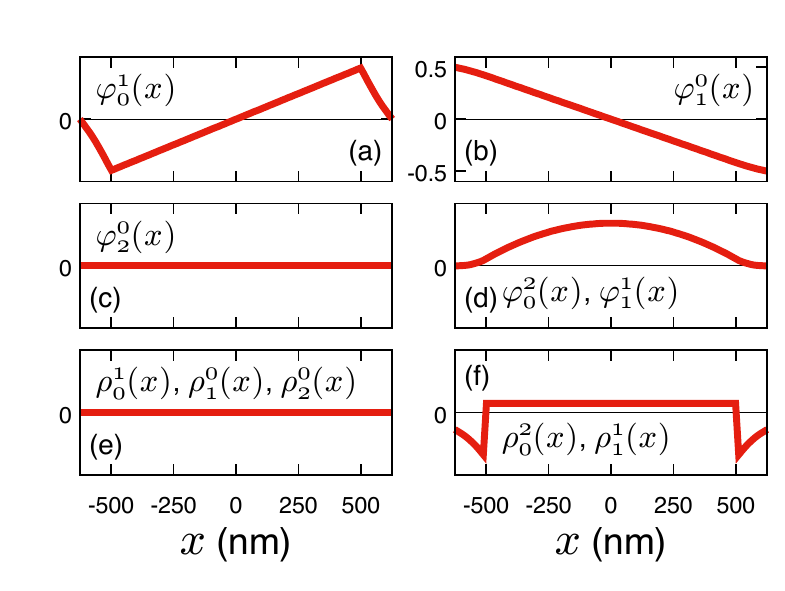}
\caption{
Analytically calculated $\varphi^n_m(x)$ and $\rho^n_m(x)$ for $n+m=1$ and $2$.
(a) The profile of $\varphi^1_0(x)$, (b) $\varphi^0_1(x)$, (c) $\varphi^0_2(x)$, (d) $\varphi^2_0(x)$ and $\varphi^1_1(x)$, (e) $\rho^1_0(x)$, $\rho^0_1(x)$, and $\rho^0_2(x)$, (f) $\rho^2_0(x)$ and $\rho^1_1(x)$.
The vertical axes are arbitrary unit except for $\varphi^0_1(x)$.
For $\rho^2_0(x)$ and $\rho^1_1(x)$, we plot $R(x)^{-2}\frac{d}{dx}\left[R(x)^2\frac{d}{dx}\right] S^2_0(x)$ because they are proportional to it.
The used parameters are the same as Fig.~\ref{fig:2}.
}
\label{fig:3}
\end{center}
\end{figure}

In Fig.~\ref{fig:3}, we plot analytically calculated $\varphi^n_m(x)$ and $\rho^n_m(x)$.
The first order term $\varphi^1_0(x)$ and $\varphi^0_1(x)$ exhibit linear slopes inside the nanochannel (Fig.~\ref{fig:3}ab), indicating that the linear response does not affect the induced charge, $\rho^n_m(x)$, (Fig.~\ref{fig:3}e).
The $(\Delta\varphi)^2$ term of $\varphi(x)$, $\varphi^0_2(x)$, is zero (Fig.~\ref{fig:3}c), indicating that the external electric field without the flux does not induce the charge inside the nanochannel.
However, the other second-order potentials, $\varphi^2_0(x)$ and $\varphi^1_1(x)$, exhibit parabolas inside the nanochannel (Fig.~\ref{fig:3}d), demonstrating that the external flow $Q$ induces the uniform electrification inside the nanochannel (Fig.~\ref{fig:3}f) because $\mathrm{Pe}$ is linear with $Q$ (see eq.~\ref{eq:Pe}). 
Although the sign and the magnitude of the induced charge due to the external flow $Q$ are complicated, the induced charge inside the nanochannel is proportional to $2-\sigma_1$ (see eqs.~\ref{eq:79} and \ref{eq:80}).  
This fact provides intuitive understanding of the nonlinear electrokinetics: the third-order electrokinetics in the nanochannel is attributed to the uniform induced charge proportional to $\mathrm{Pe}^2$ and $\mathrm{Pe}\cdot\Delta\varphi$.
In terms of the external fields $\Delta P$ and $\Delta \varphi$, the induced charge inside the nanochannel is proportional to $(\Delta P)^n(\Delta \varphi)^m$ with $n+m=2$ because $Q$ is linearly coupled with $\Delta\varphi$ and $\Delta P$ in the lowest order. 
As a result, the electrokinetic flows, $I$ and $Q$, have a nonlinear term $(\Delta P)^n(\Delta \varphi)^m$ with $n+m=3$. 

\begin{table}[t]
\caption{
Linear and nonlinear electrokinetic coefficients extracted from the previous nanochannel experiments \cite{Alice2020}.
The electrokinetic coefficients, $I^n_m$, are converted from the experimental data.
The numbers refer those described in Supplementary Information of Ref.~\citenum{Alice2020}.
The salt concentration is $\rho_0=1\,$M and $\mathrm{pH}=5.5$.
Extraction detail is described in supplementary material.
}
\label{tab:1}
\vspace{2mm}
\begin{tabular}{ccccccc}
\hline\hline
	& $I^1_0$	& $I^0_1$	& $I^3_0$	& $I^2_1$	& $I^1_2$ 	& $I^0_3$	\\\hline
	\#2	& $\sim 0$\footnote{negligibly small compared to $I^2_1$}	& 48		& $\sim 0^\mathrm{a}$ & 4000 		& $\sim 0^\mathrm{a}$ &$\sim 0$\footnote{negligibly small compared to $I^0_1$}\\
	\#3	& $\sim 0$$^\mathrm{a}$	& 32		& $\sim 0^\mathrm{a}$	& 93000 	& $\sim 0^\mathrm{a}$ & $\sim 0^\mathrm{b}$ \\
\hline\hline
\end{tabular}
\end{table}

Discussion about relevant experiments \cite{Alice2020} is crucially important for the theoretical work.
In Ref.~\citenum{Alice2020}, the measurements of the streaming current were performed in a single carbon nanotube with a radius $2\,$nm and a length about $1\,\mu$m, \cite{Alice2020} where the ionic current exhibits a quadratic dependence on the pressure difference, $(\Delta p)^2\Delta\psi$.
In their paper, this quadratic dependence was qualitatively explained by the similar electro-hydrodynamic model with the energy-barrier effect due to the confinement. \cite{Alice2020}
Here, we quantitatively discuss this experiment by the ion-wall friction. 

Table.~\ref{tab:1} lists the electrokinetic coefficients, $I^n_m$, extracted from Ref.~\citenum{Alice2020}.
The ionic current of the single carbon nanotube with varied $\Delta \psi$ and $\Delta p=0$ was perfectly linear with $\Delta \psi$ in the range of $-50\,$mV to $50\,$mV, suggesting $I^0_3$ is negligible compared to $I^0_1$. 
Furthermore, the voltage-dependent streaming current, $I_\mathrm{str}(\Delta p,\Delta\psi)=I(\Delta p, \Delta \psi) - I(0,\Delta\psi)$ exhibits the parabolic dependence on $\Delta p$, and the linear streaming current was negligible for finite $\Delta \psi$.
Therefore, $I^1_0$, $I^1_2$, $I^3_0$ are negligible compared to $I^2_1$. 
To extract the friction parameters $K$ and $\alpha_i$ from $I^1_2$, it is clear that the characteristic length at the joint between the nanochannel and reservoirs is significantly responsible for the determination of the nonlinear electrokinetics.
The importance of the joint length $\delta$ has not been recognized so far, and how long the effective $\delta$ is in the actual experiment remains unknown.   
The obtained dimensionless conductance, $I^0_1$, is in the order of $10$, suggesting that our theory yields $K\sim 10^2$ from eq.~\ref{eq.I01} because $|\sigma_0|=|\alpha_+-\alpha_-|<1$.
Furthermore, the nonlinear coefficient $I^2_1$ is in the order of $10^3$ to $10^5$, which is a significantly strong nonlinear response.   
Using eq.~\ref{eq.I21}, this is achieved only when $K\gg 1$ and $\sigma_1\neq 0$ nor $2$. 
A complete fit of all the friction parameters in the model with the experimental data is exciting, and more experiments on the water flow $Q$ are necessary.  

\section{Summary}
We have constructed the perturbation theory to calculate the nonlinear electrokinetic coefficients of confined electrolyte solution in a nanochannel.
The theory is applied to the model electro-hydrodynamic equations, which include the water-wall and ion-wall friction effects.
The linear and nonlinear electrokinetic coefficients up to the third order are analytically calculated, and simple analytical expressions for the electrokinetic coefficients are obtained in the long channel limit. 
Furthermore, the responses of the electrostatic potential profiles along the nanochannel are analytically obtained, which reveals that the uniform flow-induced charge inside the nanochannel is attributed to the ion-wall friction. 
As a result, the induced charge inside the channel has a quadratic dependence on the external fields, and thus, the third-order nonlinear electrokinetics emerges.

Our framework is general, and thus, it is easy to include non-uniform diffusion profiles under confinement.  
In this study, additional relevant effects, namely, the asymmetric geometry of the nanochannel, the fixed surface charge on the nanochannel surface, and the energy barrier of ions inside the nanochannel, are not considered within the model.
Further inclusion of these effects and quantitative comparison of the water flows and ionic currents between theory and experiments could be a fruitful frontier of future nanofluidics research. 

\section*{SUPPLEMENTARY MATERIAL}
See supplementary material for derivation of electrohydrodynamic equations via Onsager variational principle, the exact solution of expanded Poisson equations, calculation of $\Delta p^n_m$ and ${J_i}^n_m$ up to $n+m=3$ and conversion to $Q^n_m$ and $I^n_m$, and analysis on the previous experimental data.

\acknowledgments
This work was supported by JST, Presto Grant number JPMJPR21O2, JSPS Fellowships 201860001, and JSPS Grant 18KK0151, 20K14430.  
The author thanks Lyd\'eric Bocquet for the host of his stay in Paris in 2019 where he started this work and acceptance to submit it alone. 

\bibliography{nonlinear_ek}

\newpage
$\,$
\newpage

\widetext
\begin{center}
\textbf{\large Supporting Information: Analytic theory of nonlinearly coupled electrokinetics in nanochannels}
\end{center}

\setcounter{equation}{0}
\setcounter{figure}{0}
\setcounter{table}{0}
\setcounter{page}{1}
\makeatletter
\renewcommand{\theequation}{S\arabic{equation}}
\renewcommand{\thefigure}{S\arabic{figure}}
\renewcommand{\bibnumfmt}[1]{[S#1]}
\renewcommand{\citenumfont}[1]{S#1}

\section{Derivation of electrohydrodynamic equations via Onsager variational principle}
Eqs.~3 and 4 in the main article can be derived by the Onsager variational principle \cite{Doi2011_2}
 by minimizing a functional $\mathcal{R}[v_i(x),v_\mathrm{w}(x)]$ with respect to  $v_i(x)$ and $v_\mathrm{w}(x)$. 
The function $\mathcal{R}$ is defined by 
\begin{equation}
\begin{split}
	\mathcal{R}[v_i(x),v_\mathrm{w}(x)]  & = \frac{1}{2}\int\left[\sum_i\rho_i\left[\xi_i(v_i-v_\mathrm{w})^2+\lambda_iv_i^2\right] \pi R(x)^2 +\lambda_0 v_\mathrm{w}^2 2\pi R(x)\right]dx\\
& + \frac{d}{dt}\int \sum_i \rho_i\mu_i \pi R(x)^2 dx -\int p(x)\frac{d}{dx}\left[\pi R(x)^2v_\mathrm{w}\right]dx,
\end{split}
\end{equation}
where the first term is the dissipation function, the second term is the time derivative of the free energy, and the last term is the constraint of the incompressible condition, eq.~1. 

\section{The exact solution of expanded Poisson equations}
The expanded Poisson equations can be written in the form of 
\begin{equation}
\frac{1}{\kappa^2R(x)^2}\frac{d}{dx}\left[R(x)^2\frac{d}{dx}\right] \varphi^n_m(x)-\varphi^n_m(x)=-S^n_m(x),
\label{eq:ODE}
\end{equation}
Using $R(x)^2=R_0^2\mathrm{e}^{\frac{|x|-L/2}{\delta}}=R_0^2/(t+t_0)$, where $t_0=\mathrm{e}^{-\frac{H-L/2}{\delta}}$, the differential operator can be represented by
\begin{equation}
\frac{1}{\kappa^2R(x)^2}\frac{d}{dx}\left[R(x)^2\frac{d}{dx}\right]  = \left\{\begin{array}{ll}
\displaystyle \frac{(t+t_0)^2}{\kappa^2\delta^2}\frac{d^2}{dt^2} & \displaystyle \textrm{for }|x|>\frac{L}{2}, \\\\
	\displaystyle \frac{1}{\kappa^2}\frac{d^2}{dx^2} & \displaystyle \textrm{for }|x|<\frac{L}{2}.
\end{array}\right.
\end{equation}
Because eq.~\ref{eq:ODE} is an ordinary linear differential equation, the exact solution can be formally  written by
\begin{equation}
	\varphi^n_m(x) = \varphi_\mathrm{p}(x)+\varphi_\mathrm{g}(x),
\end{equation}
where $\varphi_\mathrm{p}(x)$ is the particular solution of eq.~\ref{eq:ODE}, and $\varphi_\mathrm{g}(x)$ is the general solution for the homogeneous equation. 
The general solution of the homogeneous differential equation is given by
\begin{equation}
	\varphi_\mathrm{g}(x) =\left\{\begin{array}{ll}
		\displaystyle A_+\mathrm{e}^{-\frac{k_+(|x|-L/2)}{\delta}}+A_-\mathrm{e}^{-\frac{k_-(|x|-L/2)}{\delta}}  & \displaystyle \textrm{for } |x|>\frac{L}{2},\\\\
	\displaystyle A_0\frac{\cosh(\kappa x)}{\cosh(\kappa L/2)} & \displaystyle \textrm{for } |x| < \frac{L}{2},
\end{array}\right.
\end{equation}
where $k_\pm = (1\pm\sqrt{1+4\kappa^2\delta^2})/2$ and $A_\pm$ and $A_0$ are adjustable constants.
The particular solution $\varphi_\mathrm{p}(x)$ is generally not necessary to satisfies all the boundary conditions.
The boundary conditions are $\varphi'(0)=0$, $\varphi(H)=0$, and the continuity of $\varphi(x)$ and $\varphi'(x)$ at $x=L/2$.
The adjustable constants $A_\pm$ and $A_0$ are determined to satisfies the boundary conditions, and we obtain
\begin{eqnarray}
	\varphi_\mathrm{p}(H) + A_+t_0^{k_+}+A_-t_0^{k_-} &=& 0, \\
	\Delta\varphi_\mathrm{p}(\frac{L}{2}) + A_+ + A_-  &=& A_0,\\
	\Delta\varphi_\mathrm{p}'(\frac{L}{2}) -\frac{k_+ A_+ + k_-A_-}{\delta}  &=& A_0\kappa \tanh\left(\frac{\kappa L}{2}\right),
\end{eqnarray}
where 
\begin{eqnarray}
\Delta\varphi_\mathrm{p}(\frac{L}{2}) = \varphi_\mathrm{p}(\frac{L}{2}+0)-\varphi_\mathrm{p}(\frac{L}{2}-0),\\
\Delta\varphi_\mathrm{p}'(\frac{L}{2}) = \varphi_\mathrm{p}'(\frac{L}{2}+0)-\varphi_\mathrm{p}'(\frac{L}{2}-0).
\end{eqnarray}

When the salt concentration is high enough for $\kappa L \gg 1 $ and $\kappa \delta \gg 1$, the general solution, $\varphi_\mathrm{g}(x)$ is almost zero except for near the boundary $x=\pm H$ and $\pm L/2$.
Therefore, even if $\psi_\mathrm{g}(x)$ is added to satisfy the boundary condition, $\psi_\mathrm{g}(x)$ usually does not significantly affect the charge distribution, the water flux, and the ionic currents.
Direct substitution can confirm that $\varphi^1_0(x)=S^1_0(x)$ and $\varphi^0_1(x)=S^0_1(x)$ are the particular solution of the expanded Poisson equations.
However, these particular solutions do not satisfy the continuity of ${\varphi^n_m}'(x)$ at $x=\pm L/2$.
Thus, we need to sum $\varphi_\mathrm{g}(x)$ to obtain the exact solution, but it is negligible to calculate the electrokinetic transport coefficients.

For $(n,m)=(2,0)$ and $(1,1)$, the particular solution is
\begin{equation}
\varphi_\mathrm{p}(x) =\left\{\begin{array}{ll}
	\displaystyle S^n_m(x) + \frac{A^n_m h\nu}{(\kappa^2\delta^2-2)}(t+t_0)^2 & \displaystyle \textrm{for } |x|>\frac{L}{2},\\\\
	\displaystyle S^n_m(x) - \frac{2A^n_m ht'}{\kappa^2L^2} & \displaystyle \textrm{for } |x| < \frac{L}{2},
\end{array}\right.\label{eq:phip}
\end{equation}
where the coefficient $A^n_m$ is given by 
\begin{equation}
A^n_m=\left\{\begin{array}{ll}
\displaystyle \frac{\sigma_0(2-\sigma_1)h\nu}{2} & \textrm{for }(n,m)= (2,0),\\\\
\displaystyle \frac{(2-\sigma_1)h\nu}{2} & \textrm{for }(n,m)= (1,1).
\end{array}\right.
\end{equation}
Because this particular solution does not satisfies the boundary conditions, we need to add $\varphi_\mathrm{g}(x)$ with appropriate $A_\pm$ and $A_0$. 
However, in the case of $\kappa L \gg 1 $ and $\kappa \delta \gg 1$, not only $\varphi_\mathrm{g}(x)$ but also $\varphi_\mathrm{p}(x)-S^n_m(x)$ (the second term of eq.~\ref{eq:phip}) are negligible. 
Thus, we can approximate the solutions by $\varphi^2_1(x)=S^2_1(x)$ and $\varphi^1_2(x)=S^1_2(x)$.

\section{Calculation of $\Delta p^n_m$ and ${J_i}^n_m$ up to $n+m=3$ and conversion to $Q^n_m$ and $I^n_m$}
Using the definition of $J_i$ and the approximated solution of $\varphi^n_m(x)$, we obtain ${J_i}^n_m$ up to the third-order as 
\begin{eqnarray}
{J_i}^0_0 &=& {J_i}^2_0 = {J_i}^1_1 = {J_i}^0_2 = 0,\\
{J_i}^1_0 &=& p_ih,\\
{J_i}^0_1 &=& q_ih,\\
{J_i}^3_0 & = &\frac{{p_i}^3}{4}h-3p_i h^2{f_i}^2_0(H),\\
{J_i}^2_1&=&\frac{q_i{p_i}^2}{4}h-q_ih^2{f_i}^2_0(H)-2p_ih^2{f_i}^1_1(H),\\
{J_i}^1_2&=&\frac{p_i}{4}h-2q_ih^2{f_i}^1_1(H)-p_ih^2{f_i}^0_2(H),\\
{J_i}^0_3 & = &\frac{q_i}{4}h-3q_i h^2{f_i}^0_2(H),
\end{eqnarray}
where analytical expressions for ${f_i}^2_0(H)$, ${f_i}^1_1(H)$, and ${f_i}^0_2(H)$ are as follows:
\begin{eqnarray} 
{f_i}^2_0(H) &= &\frac{q_i\sigma_0 (2-\sigma_1)}{12}h\nu t'(1+\nu t')+\frac{h(\alpha_i+\sigma_1\nu t')^2}{12}+\frac{\nu t'}{6}(\alpha_i+2\nu t')\left[\alpha_i+2\nu t'+h(\alpha_i+\sigma_1\nu t')\right],\\
{f_i}^1_1(H)&=&\frac{q_ih}{12}\left[h\left(\alpha_i+\sigma_1\nu t'\right)+\nu t'\left[(2-\sigma_1+6\alpha_i)(1+\nu t')+2\nu t'(3+4\nu t')-q_i\sigma_0 h\nu t'(3+2\nu t')\right]\right],\\
{f_i}^0_2(H)&=&\frac{1+2\nu t'}{12}.
\end{eqnarray}

The pressure difference is calculated as follows:
\begin{equation}
\begin{split}
\Delta p &= \frac{2D\eta}{R_0 b}\mathrm{Pe}+k_\mathrm{B}T\rho_0\sum_{i,n,m}F(H;\beta_i(x){\rho_i}^n_m(x))\frac{(\mathrm{Pe})^{n+1}}{n!}\frac{(\Delta\varphi)^m}{m!}-k_\mathrm{B}T\rho_0\sum_{i,n,m}p_i{J_i}^n_m\frac{(\mathrm{Pe})^n}{n!}\frac{(\Delta\varphi)^m}{m!}.
\end{split}
\end{equation}
Further calculation yields $\Delta p^n_m$ up to the third order,
\begin{eqnarray}
\Delta p^0_0& = &\Delta p^2_0 = \Delta p^1_1 = \Delta p^0_2 = 0,\\
\Delta p^1_0& = &\sum_ip_i(1-{J_i}^1_0)+\frac{2D\eta}{\rho_0 k_\mathrm{B}TR_0b},\\
\Delta p^0_1& = &-\sum_ip_i{J_i}^0_1,\\
\Delta p^3_0 &=&\sum_i \left[3F\left(H;\beta_i(x){\rho_i}^2_0(x)\right)-p_i{J_i}^3_0\right],\\
\Delta p^2_1 &=&-\sum_ip_i{J_i}^2_1,\\
\Delta p^1_2 &= & -\sum_ip_i{J_i}^1_2,\\
\Delta p^0_3 &=& -\sum_i p_i{J_i}^0_3,
\end{eqnarray}
where $F\left(H;\beta_i(x){\rho_i}^2_0(x)\right)$ are obtained as 
\begin{equation}
F\left(H;\beta_i(x){\rho_i}^2_0(x)\right) = \frac{(2-\sigma_1)h\nu t'}{12}\left[\alpha_i(\sigma_1+6\nu t')+4\nu^2t'^2\right].
\end{equation}

Obtained ${J_i}^n_m$ and $\Delta p^n_m$ are the electrokinetic coefficients in the expansion of $\mathrm{Pe}$ and $\Delta\varphi$.
It is necessary to convert ${J_i}^n_m$ and $\Delta p^n_m$ into $Q^n_m$ and $I^n_m$ in the expansion of $\Delta P$ and $\Delta\varphi$. 
$\Delta P$ as a function of $\mathrm{Pe}$ and $\Delta\varphi$ up to the third order is given by
\begin{equation}
\begin{split}
\Delta P (\mathrm{Pe},\Delta \varphi) &= \left[\Delta p^0_1\Delta \varphi + \Delta p^0_3\frac{(\Delta\varphi)^3}{6}\right]+\left[\Delta p ^1_0+\Delta p^1_2 \frac{(\Delta\varphi)^2}{2}\right]\mathrm{Pe}+\left[\Delta p^2_1\Delta\varphi\right]\frac{(\mathrm{Pe})^2}{2}+\left[\Delta p^3_0\right]\frac{(\mathrm{Pe})^3}{3!}.
\end{split}
\end{equation}
The Taylor expansion of $\mathrm{Pe}$ as a function of $\Delta P$ is given by
\begin{equation}
\mathrm{Pe}=\frac{1}{a_1}(\Delta P-a_0)-\frac{a_2}{2{a_1}^3}(\Delta P-a_0)^2+\frac{3{a_2}^2-a_1a_3}{6{a_1}^5}(\Delta P-a_0)^3,\label{eq:Pe}
\end{equation}
where
\begin{eqnarray}
a_0 &=& \Delta P|_{\mathrm{Pe}=0}=\Delta p^0_1\Delta \varphi + \Delta p^0_3\frac{(\Delta\varphi)^3}{6},\label{eq:x0}\\
a_1 &=& \left.\frac{d\Delta P}{d \mathrm{Pe}}\right|_{\mathrm{Pe}=0}=\Delta p ^1_0+\Delta p^1_2 \frac{(\Delta\varphi)^2}{2},\label{eq:a}\\
a_2 &=& \left.\frac{d^2\Delta P}{d \mathrm{Pe}^2}\right|_{\mathrm{Pe}=0}=\Delta p^2_1\Delta\varphi,\label{eq:b}\\
a_3 &=& \left.\frac{d^3\Delta P}{d \mathrm{Pe}^3}\right|_{\mathrm{Pe}=0}=\Delta p^3_0.\label{eq:c}
\end{eqnarray}
Because
\begin{equation}
\mathrm{Pe}=\sum_{n+m=0}^\infty Q^n_m\frac{(\Delta P)^n}{n!}\frac{(\Delta\varphi)^m}{m!},
\end{equation}
expansion of eq.~\ref{eq:Pe} using eqs.~\ref{eq:x0}--\ref{eq:c} gives
\begin{eqnarray}
Q^1_0&=&K \label{eq:Q10}\\
Q^0_1&=&-K\Delta p^0_1,\\
Q^3_0&=&-K^4\Delta p^3_0,\\
Q^2_1&=&-K^3\Delta p^2_1+K^4\Delta p^3_0\Delta p^0_1,\\
Q^1_2&=&-K^2\Delta p^1_2+2K^3\Delta p^2_1\Delta p^0_1-K^4\Delta p^3_0(\Delta p^0_1)^2,\\
Q^0_3&=&-K\Delta p^0_3+3K^2\Delta p^1_2\Delta p^0_1-3K^3\Delta p^2_1(\Delta p^0_1)^2+K^4\Delta p^3_0(\Delta p^0_1)^3,\label{eq:Q03}
\end{eqnarray}
where $K = 1/\Delta p^1_0$.
Therefore, using eqs.~\ref{eq:Q10}--\ref{eq:Q03}, we obtain $I^n_m$ given by
\begin{eqnarray}
I^1_0&=&\sum_i q_i{J_i}^1_0 Q^1_0,\\
I^0_1&=&\sum_i q_i\left({J_i}^1_0Q^0_1+{J_i}^0_1\right),\\
I^3_0&=&\sum_i q_i\left({J_i}^1_0Q^3_0+{J_i}^3_0(Q^1_0)^3\right),\\
I^2_1&=&\sum_i q_i\left({J_i}^1_0Q^2_1+{J_i}^2_1(Q^1_0)^2+{J_i}^3_0(Q^1_0)^2Q^0_1\right),\\
I^1_2&=&\sum_iq_i\left({J_i}^1_0Q^1_2+{J_i}^1_2Q^1_0+2{J_i}^2_1Q^1_0Q^0_1+{J_i}^3_0Q^1_0(Q^0_1)^2\right),\\
I^0_3&=&\sum_iq_i\left({J_i}^1_0Q^0_3+{J_i}^0_3+3{J_i}^1_2Q^0_1+3{J_i}^2_1(Q^0_1)^2+{J_i}^3_0(Q^0_1)^3\right).
\label{eq:I03}
\end{eqnarray}

\section{Analysis on the previous experimental data}

In Ref.~\citenum{Alice2020_2},
the ionic current across a single carbon nanotube is measured. 
The current exhibits a quadratic dependence on the applied pressure difference as
\begin{equation}
I(\Delta p, \Delta \psi) = \mu \Delta p + \left[G_0 +G_2 (\Delta p)^2\right]\Delta\psi,
\end{equation}
 where $\mu\,\Delta p$ term is negligible compared to the quadratic term $G_2(\Delta p)^2\Delta\psi$. 
The converted electrokinetic coefficients, $I^0_1$ and $I^2_1$, are obtained as follows:
\begin{eqnarray}
I^0_1 &=& \frac{k_\mathrm{B}T}{e^2\rho_0 D}\frac{L}{\pi {R_0}^2} G_0,\\
I^2_1 &=& \frac{2(k_\mathrm{B}T)^3\rho_0}{e^2 D}\frac{L}{\pi {R_0}^2}G_2,
\end{eqnarray}
where $\rho_0=1\,$M, $T=300\,$K, $D=2.0\times 10^{-9}\,$m$^2$/s, $L=1\,\mu$m, and $R_0=2\,$nm.
Table \ref{table:S2} lists the phenomenological parameters, $G_0$ and $G_2$, extracted from carbon nanotube 2 and 3 in Supplementary Information of Ref.~\citenum{Alice2020_2}, and the converted electrokinetic coefficients, $I^0_1$ and $I^2_1$.

\begin{table}[h]
\begin{center}
\caption{The electrokinetic parameters in single carbon nanotubes. \cite{Alice2020_2}}
\label{table:S2}
\begin{tabular}{>{\raggedleft}p{15mm}>{\raggedleft}p{12mm}>{\raggedleft}p{20mm}>{\raggedleft}p{12mm}r}
\hline
$\rho_0\,$(M) & $G_0$ (nS) & $G_2$ (nS/bar$^2$) & $I^0_1$ &$I^2_1$\\\hline
1 	& 4.5  	&  0.3 	& 48	& 4000	\\
1 	& 3 	&  7.0  & 32	& 93000	\\\hline
\end{tabular}
\end{center}
\end{table}

A. Marcotte, T. Mouterde, A. Nigu{\'e}s, A. Siria, and L. Bocquet, ``Mechanically activated ionic transport across single-digit carbon nanotubes,'' Nature Materials {\bf 19}, 1057-1061 (2020).

\end{document}